\documentstyle[preprint,aps,prb]{revtex}
\begin{document}

\draft
\title{Suppressed Superconductivity of the Surface Conduction Layer in
 Bi$_2$Sr$_2$CaCu$_2$O$_{8+x}$ Single Crystals Probed by {\it c}-Axis 
 Tunneling Measurements} 

\author{Nam Kim, Yong Joo Doh, Hyun-Sik Chang, and Hu-Jong Lee} 

\address{Department of Physics, 
Pohang University of Science and Technology \\
Pohang 790-784, Republic of Korea} 

\maketitle

\begin{abstract}
We fabricated small-size stacks on the surface of Bi$_2$Sr$_2$CaCu$_2$O$_{8+x}$
(BSCCO-2212) single crystals with the bulk 
transition temperature $T_c$$\simeq$90 K, 
each containing a few intrinsic Josephson
junctions. Below a critical temperature $T_c'$ ($\ll$ $T_c$),  
we have
observed a weakened Josephson coupling 
between the CuO$_2$ superconducting double layer at the crystal surface and the
adjacent one located deeper inside a stack. The quasiparticle
branch in the $IV$ data of the weakened Josephson junction (WJJ) fits well to the
tunneling characteristics of a $d$-wave superconductor($'$)/insulator/$d$-wave superconductor 
(D$'$ID) junction.
Also, the tunneling resistance in the range 
$T_c'$$<$$T$$<$$T_c$ agrees well with the tunneling in 
a normal metal/insulator/$d$-wave superconductor (NID) junction.
In spite of the suppressed 
superconductivity at
the surface layer the symmetry of the order parameter appears to remain unaffected. 
\end{abstract}

\narrowtext

\section{INTRODUCTION}
Bi$_2$Sr$_2$CaCu$_2$O$_{8+x}$ (BSCCO-2212) single crystals 
of small lateral dimensions
have been confirmed to exhibit intrinsic Josephson tunneling properties along
the {\it c} axis\cite{b1,b2,b3} due to their layered structure, where the neighboring
superconducting CuO$_2$ double layers (3 {\AA} thick) separated by 
12-{\AA}-thick non-superconducting Bi-O and Sr-O layers\cite{b1} 
form a series stack of
superconductor-insulator-superconductor tunnel junctions. Since the
{\it c}-axis coherence length for BSCCO-2212 single crystals\cite{b4} 
$\xi_c$ is only $\sim$1 {\AA}, the coupling between CuO$_2$ double 
layers is almost negligible except between the
nearest-neighbor ones so that tunneling properties of each intrinsic Josephson
junction can be well distinguished. One can fabricate stacks with junction area
as small as a few hundreds of $\mu$m$^2$ or less on the surface 
of BSCCO-2212 single crystals, each stack containing
only a few intrinsic Josephson junctions. From its current-voltage ($IV$)
characteristics one can infer the superconducting characteristics of each
conduction layer. It has been proposed, for instance, that the superconducting
order may develop in the Bi-O layer due to the proximity location to the 
CuO$_2$ double layer.\cite{b2} Also the quasiparticle 
tunneling characteristics seen in the $IV$
curves were analyzed in terms of the $d$-wave superconductivity in the
conduction layers.\cite{b3,b5} 

Recently there has been interest in the properties of the surface layer of
high-$T_c$ superconductor (HTSC) single crystals.\cite{b6,b7} The interest 
stems from
the realization that the useful tools to examine the superconducting
properties of the HTSC materials, such as the vacuum tunneling spectroscopy and the photoemission
spectroscopy,
effectively probe only the surface properties
of the crystals. On the other hand, in the 
case of transport measurements for
HTSC single crystals, one needs to deposit metallic contact 
electrodes on the
crystal surface. Aside from the possible degrading 
effect on the surface
layer such as the contamination or the deviation from the optimal
stoichiometry due to oxygen loss, deposition of a normal metal on the crystal surface
changes the boundary condition of the surface CuO$_2$ double layer along with the
corresponding changes in its physical properties. Until recently, however, few
studies have been done for the effects of normal-metal deposition on the properties of the
surface conduction layer. 

In this study, we show that {\it c}-axis tunneling measurements in stacks of a few
intrinsic tunnel junctions fabricated on the surface of HTSC single
crystals can provide valuable information on the superconducting
characteristics of the surface conduction layer in contact with the normal
electrode. From the $IV$ curves of the stacks we identified the development of
a Josephson junction right on the surface with much weakened Josephson
coupling, which was directly related to the suppressed superconductivity in the
surface conduction layer. The tunneling $IV$ curve and the {\it c}-axis resistive 
transition $R(T)$ of the
stacks supported the existence of predominant
$d_{x^2-y^2}$ symmetry in the CuO$_2$ double layers including the one
at the crystal surface. The
temperature dependence of both the critical current ($I_c'$) of this weakened
``surface junction'' and that ($I_c$) of the ``inner junctions'' embedded under the
surface junction did not follow the Ambegaokar-Baratoff relation.\cite{b8} 
$I_c'(T)$ showed a tail near the critical temperature $T_c'$ of the 
weakened Josehson junction (WJJ).
We propose that the anomalous features as well as the suppression of
the superconductivity in the surface layer resulted from the change in the
surface boundary conditions rather than material degradation of the surface
layer itself, which is believed to occur in an unprotected surface of a HTSC
material. 

\section{EXPERIMENT} 
BSCCO-2212 single crystals were grown by solid-state-reaction method.
Powders of Bi$_2$O$_3$, SrCO$_3$, CaCO$_3$, and CuO were mixed,
thoroughly ground, and heat treated in an alumina crucible.
The mixing molar ratio of Bi:Sr:Ca:Cu was 2.3:2:1:2. During
the crystal growth process, a constant oxygen gas was provided
and a temperature gradient of about 5 $^{\circ}$C/cm was set up across
the diameter of the crucible to enhance the growth rate. 
Platelets of as-grown single crystal with typical size of 
0.7$\times$0.2$\times$0.03 mm$^3$ 
were glued on MgO substrates using negative photoresist (OMR-83)
and were cleaved with Scotch-brand tape until optically smooth surfaces were
obtained. Upon cleaving a BSCCO-2212 single crystal a 500-{\AA}-thick 
layer of Au was thermally deposited on the top of the crystal to protect
the surface from contamination during further fabrication process as well as to
obtain a clean interface between the normal electrodes and the BSCCO-2212
single crystal. Stacks of size 25$\times$35$\times$0.015 $\mu$m$^3$ 
were then patterned using the conventional photolithography with positive
photoresist (Microposit 1400-23) and the Ar-ion-etching technique. Further
details of the sample fabrication are described elsewhere.\cite{b9}

Temperature dependence of the {\it c}-axis resistance $R(T)$ of a stack was measured
by conventional lock-in technique and the $IV$ data were taken using a dc
method for two different stacks \#E and \#F, which were fabricated
simultaneously on the same crystal surface. Contact configurations are shown
in Fig. 1(a). All measurements were done in a three-terminal configuration using
low pass filters connected to each measurement electrode. The bias current was
fed to the stack \#E (\#F) through the electrodes E and A (F and A) while the
voltage was taken between the electrodes E and F (F and B). Three-terminal
configuration was adopted to probe the surface junction by intentionally
including the potential drop across the surface layer. 

\section{RESULTS AND DISCUSSION} 
In a three-terminal configuration, the {\it c}-axis resistance of a stack 
consists of
the tunneling resistance of the stacked intrinsic tunnel junctions 
($R_{stack}$), both the one at the surface and the one(s) 
under the surface, and the contact
resistance ($R_c$) between the Au electrode and the BSCCO-2212 single crystal
including the resistance of the Au electrode itself. We present all of our data
below with $R_c$ ($\simeq$1.7 $\Omega$) 
subtracted, assuming that $R_c$ is temperature independent and ohmic. The
assumption is justified by the fact that the contact resistance $R_c$ 
is represented
by a linear portion below the critical current $I_c'$ in the raw $IV$ data 
(not shown)
taken at 4.2 K and by the negligible temperature dependence of $R_c$ compared
with $R_{stack}$. 
The normal-state tunneling resistivity of the stack \#E
at $\sim$100 K was
$\rho_c$=35 $\Omega$cm.

The $IV$ data of the stack \#F taken at 4.2 K [Fig. 2(a)] show a superconducting
branch and the usual multiple quasiparticle branches. Although not shown, the
stack \#E exhibited very similar feature. The multiple branches develop since
each intrinsic Josephson junction in a stack with a slightly different critical
current from one another switches separately to the resistive state with
increasing the bias current beyond $I_c$. From the number of the branches one can
easily estimate the number of the junctions (or the number of the CuO$_2$ double
layers) contained in a stack, which turns out to be 10 (11) for the stack \#F.
Each junction shows almost vertical current variation just before switching to
the resistive state. Exceeding the critical current the $IV$ characteristics of 
each
junction exhibit a characteristic voltage jump $V_c$. The value of $V_c$ upon
exceeding $I_c$ is $\sim$23 meV and, as observed previously,\cite{b1} it decreases
progressively as the number of junctions involved increases, which 
may be related to the nonequilibrium 
effect.\cite{b15} The
critical current $I_c$ in our samples turned out to be larger by a factor
than that reported previously by others,\cite{b2,b3,b5,b15} 
which may have resulted in a
larger nonequilibrium effect. All the multiple branches beyond $I_c$ exhibit large
hysteresis, which is a characteristic of an underdamped junction. 

In Fig. 2(a), we notice that the critical current of one junction is much smaller than those
of the rest of the junctions, {\it i.e.}, $I_c'$ ($\sim$0.14$I_c$), 
at which one of the 10 Josephson junctions switches to the resistive
state. This feature of the WJJ appears only below $T_c'$
[=35 K $\approx$0.39$T_c$, $T_c$ the critical temperature of the bulk 
BSCCO-2212 single crystal; see Fig. 3(a)]. 
For the samples we tested in this study, the value of $T_c'$ varied from
sample to sample in the range 20 K$\lesssim$$T_c'$$\lesssim $40 K. The possible reason for this
variation of $T_c'$ will be discussed below. The WJJ also shows
hysteresis corresponding to the McCumber parameter\cite{b10} 
$\beta_c$ of $\sim$5, which is about 100 times smaller than that of the inner 
Josephson junctions [Fig. 2(b)]. Recently, in the $IV$ curve of a stack 
fabricated on a BSCCO-2212 crystal
in the same way as used in this study, Kim {\it et al.} observed zero-current-crossing
voltage steps\cite{b11} in the resistive state of the WJJ
at voltages of 
$nh\nu/2e$ ($n$ is an integer and $\nu$
is the frequency of the rf excitation) with 
$\sim$90 GHz rf excitation. It is a strong evidence that the junction of 
the weakened
coupling forms a single Josephson junction. 

In Fig. 2(b) we compare the behavior of the quasiparticle branch of the 
WJJ with the results of numerical calculation. The $IV$ curve was
calculated using the expression
\begin{equation}
 I(V) = \frac{1}{eR_n} \int_{-\infty}^{\infty} N(E)N(E+eV)
 \left[ f(E)-f(E+eV) \right] dE
\end{equation} 
where $R_n$ is the normal-state
tunneling resistance and $f(E)$ the Fermi distribution function.
As for the quasiparticle density of states $N(E)$ we used the expression 
for the $d_{x^2-y^2}$
symmetry\cite{b3,b5,b12} as
\begin{equation}
N(E) = Re \left[ \frac{1}{2\pi} \int_{0}^{2\pi} 
\frac{E}{\sqrt{E^2-[\Delta_0 \cos(2\theta)]^2}} d\theta \right] 
\end{equation}
by modeling the surface Josephson junction as a $d$-wave
superconductor($'$)/insulator/$d$-wave superconductor (D$'$ID, D$'$ denotes an
electrode with suppressed $d$-wave superconductivity) junction with the
parameter values of 
$\Delta_0^D$=30 meV, $\Delta_0^{D'}$=($T_c'/T_c$)$\Delta_0^D$=0.39$\Delta_0^D$,
and $R_n$=2.3 $\Omega$, where $\Delta_0^D$ and $\Delta_0^{D'}$ are $\Delta_0$ for
D and D$'$, respectively [see Fig. 1(b)]. 
The value of $\Delta_0^D$ turns out to be close to the one 
of the previous reports.\cite{b3,b5} 
The fairly good fit with reasonable parameter values again strongly suggests
that the junction of the weakened coupling bears the Josephson nature with both
CuO$_2$ layers of predominant
$d_{x^2-y^2}$
symmetry. With the much smaller $I_c'$ and the lower $T_c'$ than the other intrinsic
Josephson junctions, the superconducting order parameter of one of the 11
CuO$_2$ double layers must be significantly suppressed with the resultant weakened
interlayer coupling across the corresponding junction(s). Since the CuO$_2$
double layer of the suppressed superconductivity affects the characteristics of
only one junction instead of two the layer must be the topmost one located
closest to the stack surface. The location of the WJJ on the surface
has also been ascertained by measurements using {\it in situ} thickness control 
of a
stack by Doh and Lee.\cite{b9} The authors observed that the tunneling 
characteristics
of the WJJ appeared prior to the appearance of any other multiple
branches in the $IV$ characteristics. 

As to the cause of the order parameter suppression, we pay attention to the fact
that the CuO$_2$ double layer at the crystal surface is in contact with the 
normal
Au electrode, while the inner CuO$_2$ double layers are Josephson coupled to the
neighboring ones. Deutcher and M\"{u}ller\cite{b6} predicted suppression of the pair
potential of a HTSC with a short {\it c}-axis coherence length $\xi_c$
when in contact with a non-superconducting layer perpendicular to
it.\cite{b20}
This proximity effect of the normal metal on the
superconductivity of HTSC are not well understood to date. Although not
directly applicable to HTSC systems, the McMillan model\cite{b13} for conventional
superconductors predicts that the superconducting transition temperature $T_{NS}$
of the S electrode in a normal-metal/superconductor (NS) junction 
depends on the thickness of each metal and the
electron scattering strength at the interface between N and S. The model
predicts that the thinner the S layer is in comparison with
the N layer, the less stable the superconductivity becomes in the
S layer. Since the CuO$_2$ double layer is extremely thinner than
the Au electrode, we may assume that the pair potential in the surface
CuO$_2$ layer was significantly suppressed. 

Fig. 3(a) shows the temperature dependence of the {\it c}-axis tunneling resistance
$R(T)$ for the stacks \#E and \#F with the measurement configuration as
illustrated in Fig. 1(a). Although the stacks \#E and \#F were fabricated
simultaneously on the same crystal surface, the resistance of the stack \#F,
just above $T_c$ ($\simeq$90 K), 
is about a factor of two larger than that of the stack \#E,
presumably because the voltage for the stack \#F was across more junctions
[including the ones in the larger base as illustrated in Fig. 1(a)] than for the 
stack \#E.
Nonetheless, the $R(T)$ data below $T_c$ for both stacks \#E and \#F 
merge to a single
curve, showing a single peak at $T_c'$ ($\simeq$35 K) 
and an ensuing abrupt superconducting transition. This suggests again
that the resistive curve in the range $T_c'<T<T_c$ is a surface effect, 
the same one
as adopted for the analysis of the $IV$ curves. In the intermediate temperature
range $T_c'<T<T_c$, the inner CuO$_2$ double layers below the surface conducting
layer are superconducting while the surface CuO$_2$ double layer remains normal
due to its suppressed superconductivity [Fig. 1(b)]. Thus, as far as the low-bias resistive
state is concerned, each stack in this temperature range behaves like a single
NID junction (N denotes a normal electrode). Due to the thermally assisted
nature of the quasiparticle tunneling from N to D, the tunneling resistance of
the surface junction increases with decreasing temperature until the weakly
coupled Josephson junction forms at $T_c'$. Assuming $N(E+eV)$=1 in Eq. (1) along with
the BCS-type temperature dependence of the gap $\Delta_0$ in Eq. (2),
one can
calculate the temperature dependence of the tunneling resistance $R$ for a NID
junction from the relation $R=(dI/dV)^{-1}|_{V=0}$ using the same value of the 
gap
energy ($\Delta_0^D$=30 meV) as used for the $IV$-curve calculations in Fig. 2(b). 
We used the
normal-state tunneling resistance $R_n$ (=1.55 $\Omega$)
as the fitting parameter, which is reasonably close to the value 
$R_n$=2.3 $\Omega$
used in the $IV$-curve calculations [Fig. 2(b)]. Below $T_c$, the calculated 
$R(T)$
curve (dotted line) shows almost the same curvature and fits rather well to the
measured $R(T)$ data of both stacks. Also, illustrated in Fig. 3(b) is the
gradual change of the $IV$ characteristics of the WJJ with
temperature. One clearly notices that a D$'$ID-like behavior of the surface
junction with hysteresis gradually transforms into a monotonous NID-like
behavior with a finite subgap states in D as the temperature approaches $T_c'$
from below [refer to Fig. 1(b)]. 

Since the weakened Josephson coupling is considered to be a phenomenon related to the proximity
coupling between the superconducting surface CuO$_2$ double layer and the
normal-metallic electrode, it should be absent for a four-terminal configuration
(where the CuO$_2$ layers at the surface under the current and voltage contacts are separated)
or in a stack with heat-treated contact electrodes. In the latter case no sharp
boundary between the normal electrode and the CuO$_2$ double layer can be defined
due to the thermal interdiffusion of the composition materials. 
Yurgens {\it et al.}\cite{b14}
also reported, without interpretation, similar secondary resistance peak in $R(T)$
like ours far below $T_c$ for a three-terminal measurement configuration, while
no such feature was observed by the authors in a four-terminal configuration.
No feature of the WJJ was reported from other
experiments,\cite{b2,b15} where either four-terminal configuration or heat-treated
electrodes were used. The typical feature of the WJJ was
observed in all the samples we examined, when the data were taken in a
three-terminal configuration, although $T_c'$ and $I_c'$ may differ from 
sample to
sample. Currently we have no clear explanation for the relatively large
variation of $T_c'$ and $I_c'$ for different samples, while the values of 
$T_c$ were 
reproducible. As mentioned above, however, since the $T_{NS}$ of S in a NS
junction is sensitive to the scattering barrier or the contact resistance of the
junction, we suppose that the lack of reproducibility of $T_c'$ and $I_c'$ 
may be due to
the sample-to-sample difference in $R_c$. The coupling between two adjacent
Bi-O layers is known to be weakest in the BSCCO-2212 crystal. Thus, upon
cleaving, a Bi-O layer is most likely to be surfaced. However, the influence of
the existence of this layer on the suppression of the order parameter in the
surface CuO$_2$ double layer is currently unclear and is a subject of 
further study. 

We have also obtained the temperature dependence of the critical current
$I_c'(T)$ and the return current $I_r'(T)$ of the WJJ (Fig. 4) from
the measured $IV$ characteristics. In the low temperature region the values 
of $I_c'$
have rather high fluctuations, which may be attributed to the multi-valued
critical current induced by fluxon motions in the long junction limit\cite{b16}
[junction width$>$Josephson penetration depth ($\lambda_J$ $\sim$1 $\mu$m) for BSCCO-2212].
The $I_c'(T)$ curves for both stacks \#E and \#F strongly deviate from the
Ambegaokar-Baratoff expression and exhibit a long tail in the range 
$\frac{1}{2}$$T_c' \lesssim T \lesssim T_c'$. On the other hand, the critical current 
$I_c(T)$ of the inner Josephson
junctions does not show the tail structure (see the inset of Fig. 4). Curiously
enough, $I_c(T)$ appears to follow the temperature dependence of the BCS
superconducting gap. The tail structure of $I_c'(T)$ cannot be explained by the
present S$'$IS- or D$'$ID-type single junction model\cite{b10,b19} (S$'$ denotes a
superconducting electrode with suppressed $s$-wave superconductivity). This anomalous
temperature dependence of the WJJ in comparison with
the inner Josephson junctions may provide the clues to the origin of the
suppressed superconductivity. Studies on the temperature dependence of the
critical current have been done, both theoretically and experimentally, for
NSIS-type proximity junctions consisting of conventional
superconductors.\cite{b17,b18} In an experimental work\cite{b18} 
with conventional
superconductors Camerlingo {\it et al.} observed a small tail in the 
critical current
near the transition temperature, although theoretical calculations\cite{b17} did not
predict the tail structure. 

In conclusion, in stacks of intrinsic Josephson junctions formed on the surface
of BSCCO-2212 single crystals, we have observed a WJJ
presumably between the surface superconducting layer and the
adjacent inner superconducting layer at temperatures below $T_c'$=35 K, significantly lower than
the $T_c$ of the bulk BSCCO-2212 single crystals. The quasiparticle branch in the
$IV$ data of the surface Josephson junction fits well to the tunneling behavior of
the assumed D$'$ID-type junction. We could also explain the increase in its
tunneling resistance with decreasing temperature below $T_c$, assuming the
surface junction as a NID junction. The proximity location of the surface
conduction layer to the normal electrode suppresses the superconductivity in
the surface layer. Nonetheless, apparently it does not alter the symmetry of the
suppressed superconductivity, the predominant 
$d_{x^2-y^2}$
nature remains intact. Despite the qualitative agreements between some
experimental data and the single junction model, we still need more
comprehensive theoretical treatment for the superconducting characteristics
existing at the surface of high-$T_c$ superconductors. Although the suppression
of the superconducting order in the surface CuO$_2$ layer is certainly an
unfavorable phenonmenon for any device application it may still be positively
utilized to study the microscopic properties of the inner CuO$_2$
layer such as the quasiparticle density of states using the 
{\it c}-axis NID tunnel-junction configuration. It
can also be exploited to study the order parameter symmetry of HTSC
materials by forming a conventional-superconductor/high-$T_c-$superconductor
{\it c}-axis hybrid junction, since the symmetry of the surface layer does not change from
that of the bulk. 

\section{ACKNOWLEDGMENTS}
This work was supported by the Korea Science and Engineering Foundation
under Contract No. 1NK9700501, BSRI administrated by the Ministry of Education
under Contract No. 1RB9811401, and the Ministry of Defense through MARC
under Contract No. 1MC9801301.

\begin{figure}
\caption{ (a) Schematic diagram of the stacks fabricated on a BSCCO-2212 single
crystal. The size of the stacks \#E and \#F is 
25$\times$35$\times$0.015 $\mu$m$^3$. 
Measurement configuration of the current and the voltage is as follows: 
$I$(E$\rightarrow$A), $V$(E$\rightarrow$F) for the stack \#E; 
$I$(F$\rightarrow$A), $V$(F$\rightarrow$B) for the stack \#F.
(b) Schematic structure of a stack near the surfacce
in the temperature ranges $T$$<$$T_c'$ and $T_c'$$<$$T$$<T_c$,
respectively. 
As discussed in the text the Au electrode and the surface CuO$_2$
double layer form an ohmic contact.
The insulating barrier (I) is believed to form
between neighboring Bi-O layers.}
\end{figure}

\begin{figure}
\caption{ $IV$ curves for the stack \#F measured at 4.2 K are plotted with 
the contact
resistance subtracted. (a) Multiple quasiparticle branches with the critical
current of the surface junction ($I_c'$) and that of the inner junction ($I_c$). 
(b) The
$IV$ curve of the surface junction (solid curve) showing a hysteresis. 
The dotted
curve shows the calculated $IV$ curve for a D$'$ID junction using 
Eqs. (1) and (2)
with $\Delta_0^D$=30 meV, $\Delta_0^{D'}$=11.6 meV, and $R_n$=2.3 $\Omega$. }
\end{figure}

\begin{figure}
\caption{ (a) The $R(T)$ curve of the stacks \#E (circle) and \#F (triangle) 
for a bias
current $I_b$=0.5 $\mu$A with the contact resistance 1.72 $\Omega$ 
subtracted. Dotted curve exhibits the calculated fit with 
$\Delta_0^D$=30 meV and $R_n$=1.55 $\Omega$.
(b) The $IV$ curves of the WJJ for the stack \#E at various temperatures. The curves are
displaced horizontally for clarity. As the temperature approaches $T_c'$ from
below the critical current as well as the hysteresis gradually disappear.  }
\end{figure}

\begin{figure}
\caption{ Temperature dependence of critical current of the surface Josephson
junction $I_c'(T)$ of the stacks \#E (solid circle) and \#F (open circle), and the
return current $I_r'$ (dotted line) of the stack \#E. 
Inset: temperature dependence of the
critical current of an inner junction $I_c(T)$ of the stack \#E (solid circle). 
The
BCS gap relation (solid line) and the AB relation (dotted line) are plotted
together. }
\end{figure}


\begin{references}
\bibitem{b1} R. Kleiner, F. Steinmeyer, G. Kunkel, and P. M\"{u}ller, 
Phys. Rev. Lett. {\bf 68},
2394(1992); R. Kleiner and P. M\"{u}ller, Phys. Rev. B {\bf 49}, 1327 (1994). 
\bibitem{b2} A. Yurgens, D. Winkler, N. V. Zavaritsky, and T. Claeson, 
Phys. Rev. B {\bf 53}, R8887 (1996). 
\bibitem{b3} K. Schlenga, R. Kleiner, G.Hechtfischer, M. M\"{o}{\ss}le, 
S. Schmitt, Paul M\"{u}ller, Ch. Helm, Ch. Preis, F. Forsthofer, J. Keller, 
H. L. Johnson, M. Veith, and E. Steinbei{\ss}, Phys. Rev. B {\bf 57}, 1 (1998). 
\bibitem{b4} T. T. M. Palstra, B. B\`{a}tlogg, L. F. Schneemayer, 
R. B. van Dover, and J. V. Waszczak, Phys. Rev. B {\bf 38}, 5102 (1988); 
J. H. Kang, R. T. Kampwirth, and K. E. Gray, 
Appl. Phys. Lett. {\bf 52}, 2080 (1988). 
\bibitem{b5} M. Itoh, S. Karimoto, K. Namekawa, and M. Suzuki, 
Phys. Rev. B {\bf 55}, R12001 (1997). 
\bibitem{b6} G. Deutscher and K. A. M\"{u}ller, Phys. Rev. Lett. {\bf 59}, 1745 (1987). 
\bibitem{b7} T. Giamarchi, M. T. B\'{e}al-Monod, and Oriol T. Valls, 
Phys. Rev. B {\bf 41}, 11033 (1990); 
S. H. Liu, and R. A. Klemm, Phys. Rev. B {\bf 52}, 9657 (1995); 
Stephen W. Pierson, and Oriol T. Valls, Phys. Rev. B {\bf 45}, 2458 (1992); 
R. S. Gonnelli, D. Puttero, G. A. Ummarino, V. A. Stepanov, and F. Licci, 
Phys. Rev. B {\bf 51}, 12782 (1995). 
\bibitem{b8} V. Ambegaokar and A. Baratoff, Phys. Rev. Lett. {\bf 10}, 
486 (1963); {\bf 11}, 104(E) (1963). 
\bibitem{b9} Yong Joo Doh and Hu-Jong Lee, {\it Proceedings of the 8th Conference on
Materials and Applications of Superconductivity, Yong-Pyung, Korea, 1998}. 
\bibitem{b15} K. Tanabe, Y. Hidaka, S. Karimoto, and M. Suzuki, 
Phys. Rev. B {\bf 53}, 9348 (1996). 
\bibitem{b10} A. Barone, {\it Physics and Applications of the Josephson Effect} 
(John Wiley \& Sons, Inc. 1982). 
\bibitem{b11} Jinhee Kim, Kyu-Tae Kim, Se Il Park, Yong Joo Doh, Nam Kim, and
Hu-Jong Lee (unpublished). 
\bibitem{b12} H. Won and K. Maki, Phys. Rev. B {\bf 49}, 1397 (1994). 
\bibitem{b20} In the theory, however, $d$-wave symmetry was not assumed in the HTSCs. The
relevance to the HTSCs was introduced only in terms of extremely short {\it c}-axis
coherence length.
\bibitem{b13} W. L. McMillan, Phys. Rev. {\bf 175}, 537 (1968). 
\bibitem{b14} A. Yurgens, D. Winkler, N. V. Zavaritsky, and T. Claeson, 
Proc. SPIE {\bf 2697}, 433 (1996). 
\bibitem{b16} N. Mros, V. M. Krasnov, A. Yurgens, D. Winkler, and T. Claeson, 
Phys. Rev. B {\bf 57}, R8135 (1998). 
\bibitem{b19} Yukio Tanaka and Satoshi Kashiwaya, 
Phys. Rev. B {\bf 56}, 892 (1997).
\bibitem{b17} V. Z. Kresin, in {\it Josephson Effect: Achievements and Trends}, 
edited by A. Barone (World Scientific, Singapore,1985), p.198. 
\bibitem{b18} C. Camerlingo, R. Monaco, B. Ruggiero, M. Russo, and G. Testa, 
Phys. Rev. B {\bf 51}, 6493 (1995). 
\end{references}
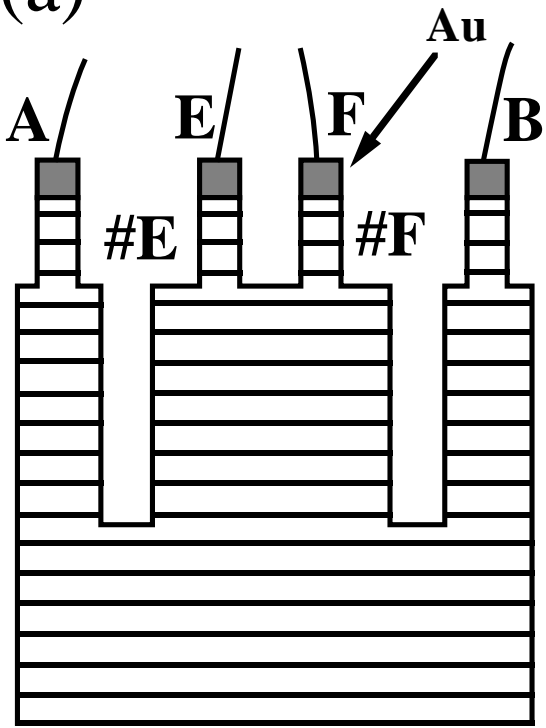
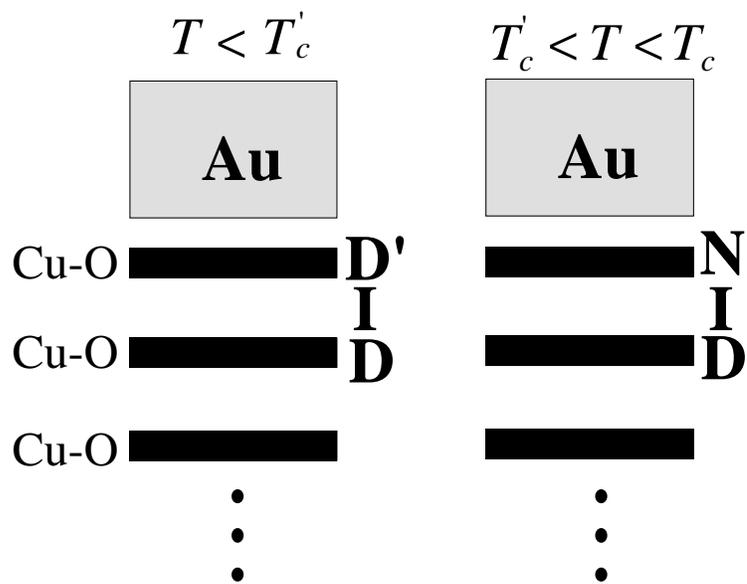
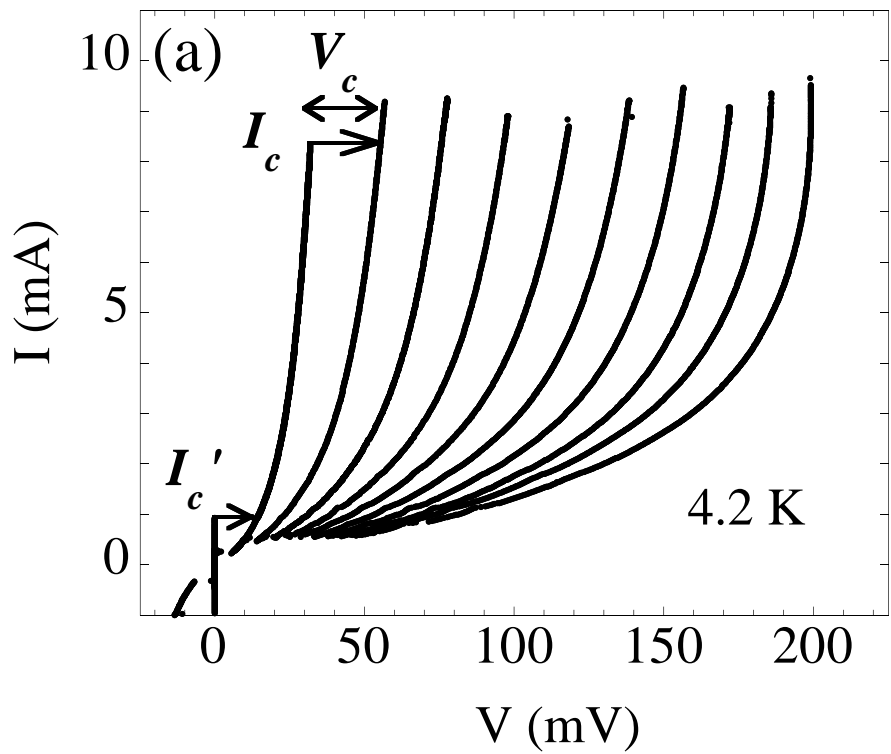
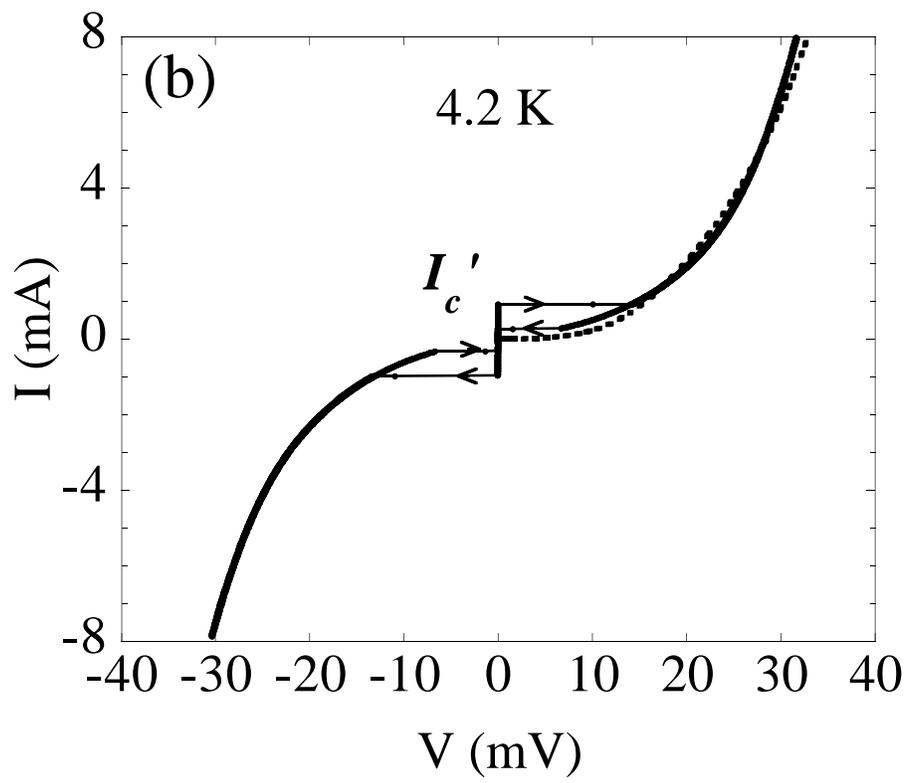
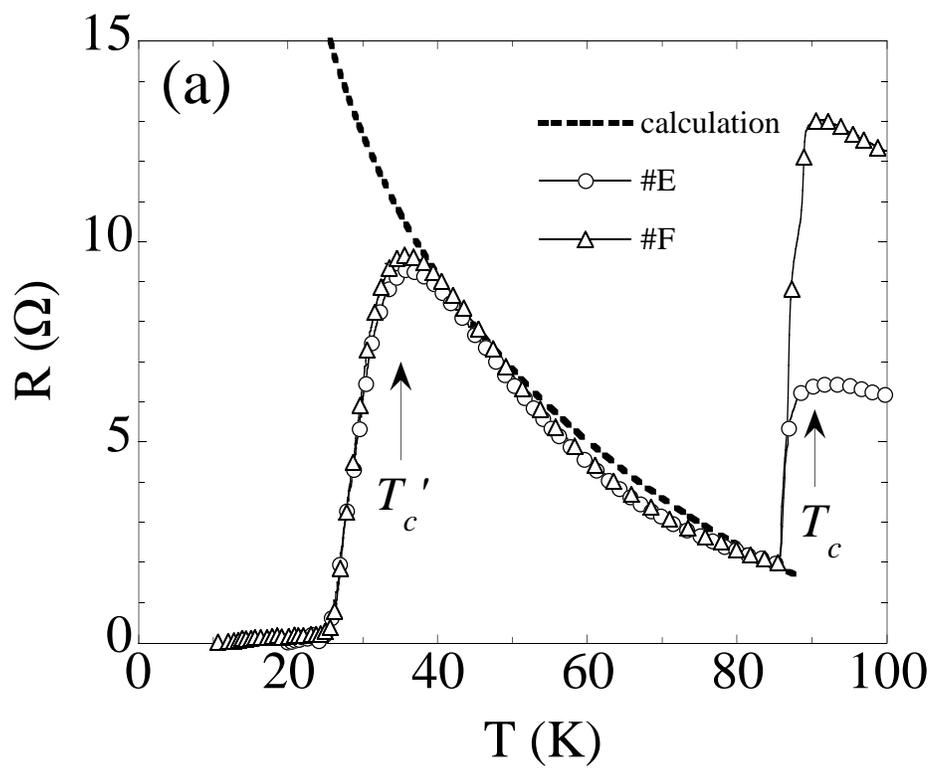
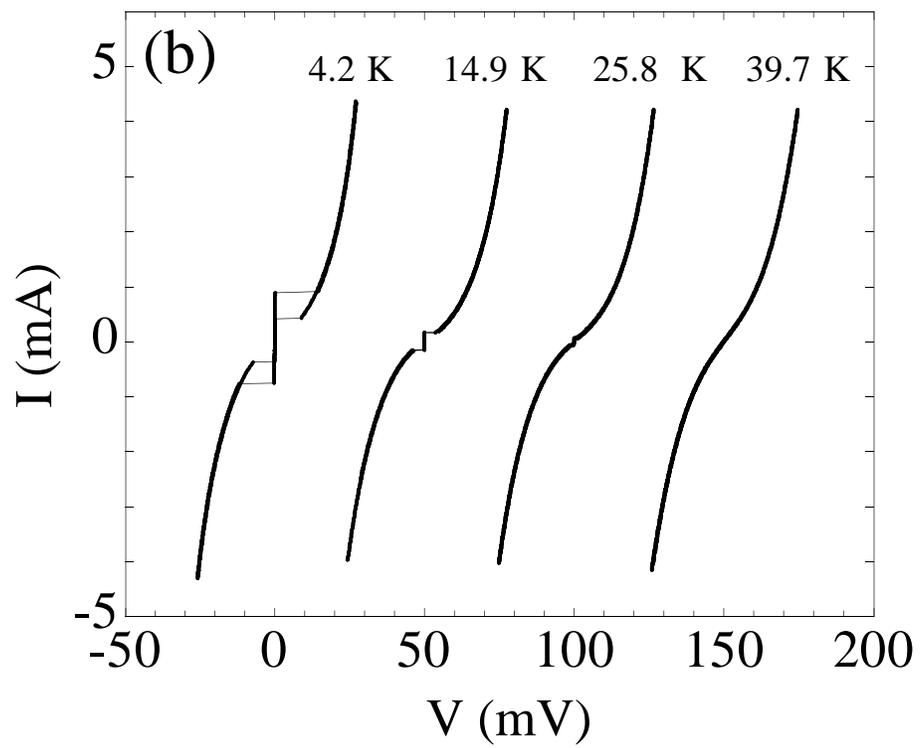
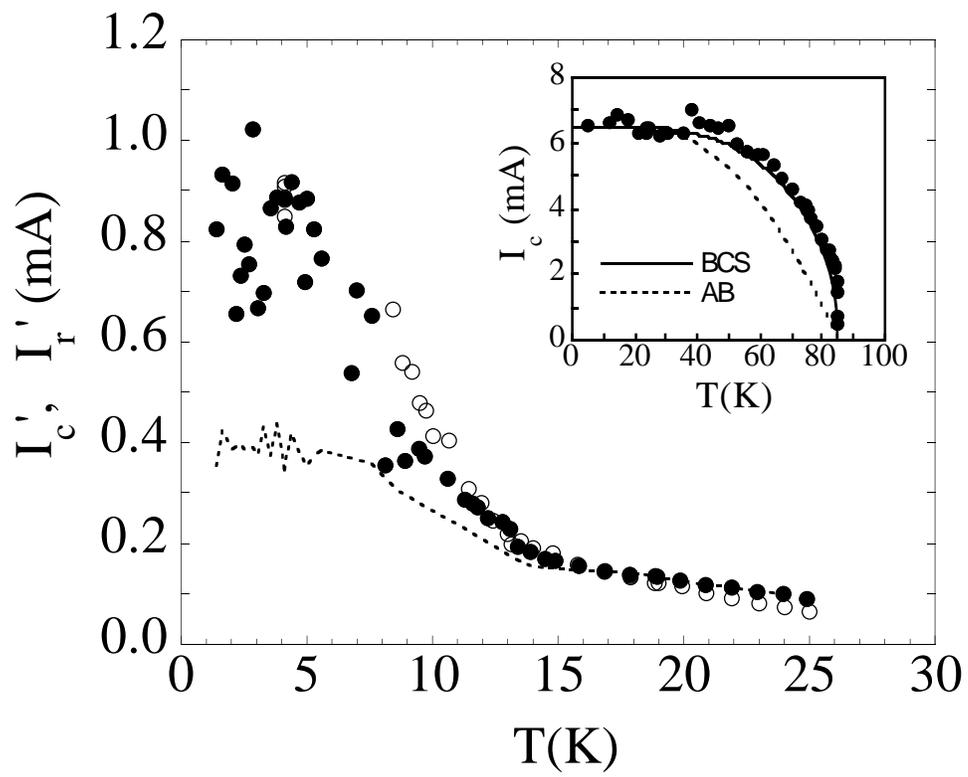
\end{document}